\documentclass[12pt,a4paper]{article}
\usepackage[dvips]{graphicx}
\usepackage{epsfig}
\begin{document}
\begin{center}
{\bf\Large Theoretical Investigation of One-Dimensional Cavities in Two-Dimensional Photonic Crystals}\\
\vspace{10mm}
\end{center}
\begin{flushleft}
\vspace{10mm}
{\large S. Foteinopoulou $^{1}$ and C.M. Soukoulis $^{1,2}$}\\
\end{flushleft}
\begin{flushleft}
{\it $^1$Ames Laboratory-USDOE and Department of Physics and Astronomy \\
Iowa State University, Ames, Iowa 50011} \\
\end{flushleft}
\begin{flushleft}
{\it $^2$Research Center of Crete-FORTH and Department of Physics,\\University of Crete, Heraklion, Crete 71110, Greece}\\
\end{flushleft}

\vspace{1cm}

{\small We study numerically the features of the resonant peak of one-dimensional (1-D) dielectric cavities in
a two-dimensional (2-D) hexagonal lattice. We use both the transfer matrix method and the finite
difference time-domain (FDTD) method to calculate the transmission coefficient. We compare the two methods and discuss their results for the transmission and quality factor Q of the resonant peak. We also examine the dependence of Q on absorption and losses, the thickness of the sample and the lateral width of the cavity. The Q- factor dependence on the width of the source in the FDTD calculations is also given.} 

\newpage 
{\bf\large I. Introduction}
\vspace{0.5cm}

In semiconductor crystals the presence of the periodic potential
affects the properties of the electrons. Likewise in photonic 
crystals (PCs), that are periodic dielectric arrangements in one, two or 
three dimensions the properties
of the photon can be controlled. For certain frequency regions known as
``the photonic band gap'', propagation of electromagnetic (EM) waves is
prohibited in the PC \cite{Joannp,sedbookp,sedbook2,Villenp}.
A defect present in an otherwise periodic crystal may introduce one or
more propagation modes in the band gap and forms
a cavity. Cavities of different geometries can exist in a photonic crystal
and the resulting resonant mode can be classified into two general types indicating
whether it drops from the ``air'' (higher) band or raises from the ``dielectric''
(lower) band \cite{sedp1}. The first type is associated with defects corresponding to the removal
of dielectric material while the second corresponds to defects involving
the addition of dielectric material. The features of the resonant mode
will depend on both the bulk crystal and the cavity characteristics and can so be tuned 
accordingly \cite{Joannp,sedbook2,sedp1}. 

An important virtue of the PC cavity is that it can control 
the 
atomic spontaneous emission, while metallic cavities for the related frequency range are
generally lossy \cite{LeeJOSAB629,sedoldp355,SLAPL3233}. Spontaneous emission is important
for a
number of semiconductor devices \cite{yabJOMO173}. Resonant cavities have undergone 
extensive studies \cite{SLAPL3233,SmithJOSAB2043,SmithAPL1487,HwangJOSAB2316,SigalasPRB3815}, both theoretical and experimental. A variety of applications incorporating
a PC cavity have been  
suggested or reported \cite{HirayAPL791,Burak,Burak2}, such as optical
laser components \cite{Joannp,sedbookp,sedbook2}, optical filters \cite{sedp1},
single mode light emiting diode (SMLED) \cite{yabJOMO173,sedoldp369} and optical imaging \cite
{SievenAOPT2074}.

In this paper we focus on the theoretical study of an 1-D dielectric cavity in between a hexagonal
patterned region of air cylinders in a dielectric matrix for the case of H polarization (magnetic field along the cylinder axis). We will present the cavity mode frequency 
versus the cavity length ($L_{c}$) (see Fig. 1), as well as the quality factor (Q) of the modes versus their 
corresponding frequency. The quality factor is defined as $ \lambda_{p}$/$ \delta$$ \lambda$ where $ \lambda_{p}$ the frequency of the resonant peak and $ \delta\lambda$ the width of the resonance at transmission half of its maximum value. In our case the quality factor Q of the resonance is relatively small ($\sim$200), and other ways of determining the Q factor (such as from the energy decay of the resonance \cite{Agio}) give similar results. The dependence of the resonant mode on the size of the system (in the lateral direction and in the direction of propagation) will be
also shown. In our calculations, two different numerical techinques (the transfer matrix and the FDTD) are used and
the results of the two methods are
compared and discussed. With both techniques we calculate the transmission through 
the structure and from that we
obtain all the relevant features of the resonant mode. 

In Section II we describe briefly the two calculational methods. In Section III we present and discuss our calculational results concerning the quality factor and position of the cavity resonance. We also compare the latter with the values obtained from the experiment previously performed on the structure under study \cite{rattier}. In Section IV we give a summary of our results.\\

\vspace{1cm}
{\bf\large II. Calculational methods}
\vspace{0.5cm}

The first technique we use is the transfer matrix method (TMM) developed by Pendry
\cite{Pendry1,Pendry2}. In the TMM
a grid lattice is used to discretize the space, and the structure is divided into finite blocks
along the propagation direction. With the use of the Maxwells equations,
 that are solved
on the grid lattice, the 
electric and magnetic field can be integrated throughout the blocks and so the respective 
transmission coefficient can be calculated. 
The final transmission will result by combining the ones
of the individual blocks. In the TMM the modeled structure is finite in the
propagation direction (y) but infinite in the x,z directions (where z is the direction of the air cylinders),
and the incoming electromagnetic wave is a plane wave. The structure is embedded in a medium
with dielectric constant equal to the background dielectric constant to simulate the experiment \cite{rattier}.

The second method we use is the finite difference time domain (FDTD) technique \cite{Agio,Tavlove}. The real space is discretized on a grid lattice and Maxwell's equations are solved in time domain. The electric and magnetic fields are updated in every point of the grid lattice in finite time steps. The structure is infinite in the direction of the cylinders (z direction) but has finite size in both x and y direction and is embedded in a finite sized (in x and y) dielectric slab with dielectric constant equal to that of the matrix medium. Liao absorbing boundary conditions \cite{liao} are applied at the walls of the slab to avoid reflections. The source is a pulse, located close to x=0, y=0 ( y being the direction of propagation), has finite length and generates fields with a trapezoidal (extented with gaussian tails) profile in space. To calculate the transmission, a line detector is placed along the latteral direcion right after the structure. The component that is perpendicular to the detector of the Poynting vector (for the Fourier transformed fields) is taken and averaged over the detector. This is normalized with the respective value for the same detector but positioned close to the source in the absense of the structure (to avoid corrupting the data with reflections) and yields the transmission of the structure. 

In both methods the quality factor is calculated from the data of transmission versus frequency data from $ \lambda_{p}$/$ \delta$$ \lambda$ where $ \lambda_{p}$ the frequency of the resonant peak and $ \delta\lambda$ the width of the resonance at transmission half of its maximum.
 
The two methods differ not only in the calculational approach but also in the characteristics of the source of the incoming EM waves, and in the lateral size of the sample that is simulated. We also performed calculations using the FDTD method but applying Bloch (periodic) boundary conditions in the lateral (x) direction to have conditions similar to those for  the transfer matrix. The source that is used in the latter case is still a pulse but with a plane-wave front.

\vspace{1cm} 
{\bf\large III. Results}
\vspace{0.5cm}

In Fig. 1 the cross section of the cavity structure under study with the x-y plane can be seen. It corresponds to air cylinders ($\epsilon$$_{a}$= 1.0) in a GaAs background ($\epsilon$$_{b}$= 11.3) (the value of the GaAs dieletric constant for $\sim$1000 nm is taken in the calculations for simplicity). The whole cavity structure is embedded in GaAs ($\epsilon$= 11.3). The symmetry directions of the bulk crystal are also shown. With $L_{c}$, we will refer to the cavity width that corresponds to the length of the dielectric defect introduced along the propagation direction. With $L_{w}$, we will refer to the lateral width of the cavity (which is the same as the sample's lateral width). $N_{c}$ will be the number of rows in each side of the cavity ($N_{c}=4$ for both calculations and experiment). The thickness of the sample then along the propagation direction would be $L_{th}=2 N_{c} b+ L_{c}$, where 2b=$\sqrt3$a ( but is approximated with 1.7a in the TMM for numerical reasons). (Such an approximation introduces  an error of at most 3$\%$ in the position of the resonance and does not affect the quality factor). The radius R of the air holes is R$\sim$0.2803a  (a being the lattice constant) that corresponds to a filling factor of $\sim$0.285. (Actually in the TMM because of the approximation mentioned above the simulated structure will have a slightly larger filling factor). No specific value for the lattice constant a is assumed. All the band gap and transmission properties in a photonic crystal scale with the lattice constant a, so all the subsequent results will be presented in dimensionless units of frequency (reduced frequency u=a/$\lambda$) and length.   

From the transfer matrix results for the various $ L_{c}$ we observed that the spectral gap when the cavity is introduced is wider than the one for the periodic crystal. Such widening of the gap was also observed in \cite{SigalasPRB3815} when a defect is introduced by removal of 2 rows of sites in a periodic 2-D system of dielectric rods in air background. TMM calculation have been performed for cavity structures with dimensionless $L_{c}/a$ ranging from $\sim$0.15 to 2.0. At a certain frequency different resonances (peaks) can occur corresponding to different $L_{c}$ values. Each of these peaks is characterized by an order, that basically indicates the order at which a resonant peak appears at this specific frequency while $L_{c}$ is increasing. 
 
In Fig. 2 every cavity width (in units of a) is plotted as a function of the corresponding reduced frequency of the observed peak and three different curves are recovered for the three orders of the resonant peaks. Starting from the lower to the higher curve in Fig. 2 the peaks are the first-, second- and third- order, respectively. It will be seen later that peaks of the same frequency but different order can have different features (e.g., quality factor). It is evident from Fig. 2 that for certain $L_{c}$ values more than one resonant peaks can appear within the gap.    
The experimental results \cite{rattier} for the resonant peaks for various lattice constants (a=200 nm, 210nm, 220nm, 230nm, 240nm) are also included in the figure.
The agreement is generally good but there seems to be a small discrepancy that increases with the frequency. This can be attributed to the fact that we have taken the GaAs dielectric constant not to vary with the frequency (for simplicity the value at $\sim$1000 nm is taken in our calculations) while this is not the case in the actual system. This difference in the dielectric will alter slightly the position of the peaks \cite{rattier}. This is consistent with the fact that the discrepancy at the higher end of the reduced frequencies seems to be a little larger for the smaller lattice constant (the corresponding wavelength differs more from the value of a 1000 nm). There are also some experimental peaks for a=200 nm that are falling above the theoretical curves, as for example, the first order peaks with $L_{c}$/a from $\sim$0.60 to 0.85.  We have not numerically calculated the Qs for these, because they lie too close to the band edges and the determination of their quality factor can be vaque and so we have not included them in Fig. 2.    
 
We compare now the two computational methods ( TMM and FDTD) that are used to simulate the cavity structure. 
There are basically three fundamental differences between the two methods. 
The first one is that the transfer matrix technique is a time independent method while
 in the FDTD the fields are solved in time domain. 
Also, in the TMM, the structure is infinite in x by virtue of the periodic boundary 
conditions applied along this direction, 
while in the FDTD the system is finite in the x-y plane bounded by 
absorbing boundary conditions. Finally, the incoming EM fields in the TMM framework are extended
plane waves incident
on the whole x-z boundary (corresponding to a source infinite in width) while in the FDTD a pulse emitted from a source with finite width 
is considered. Also the way the transmission coefficient is obtained is different in the two methods. In the TMM it is calculated from the transfer matrix using the field values while in the FDTD it is calculated from the energy ratio that passes through the crystal along the propagation direction. In order to make the comparison between the two methods in this particular FDTD calculation, the 2-D space is discretized as in transfer matrix 
which gives 2b=1.7a
instead of $\sqrt3$a . The results for one particular cavity width are shown in Fig. 3. Good
agreement is found between the two methods regarding the positions of the resonant peak and the
band gap edges but there is a small discrepancy in the quality factor of the peak.

It is interesting therefore to see how the quality factor of the resonant peak is affected by
the finiteness of the size of the system and the width of the source, in our FDTD calculations. This can be seen in Fig. 4(a) and 4(b) where
the quality factor versus the lateral width of the sample ($L_{w}$) and versus the source width are 
shown respectively. In Fig. 4(a) the ratio of the source width over the lateral extent of the structure $L_{w}$ (defined in Fig. 1) is kept
constant (equal to 0.5). 
It can be seen that as the lateral width approaches 20a, Q saturates at a value$\sim$170. In Fig. 4(b) we present results of the Q versus the source width for a system with a lateral width of $L_{w}=40a$ . The width $L_{w}$ is large enough for the quality
factor to have reached it's saturation value (which is related to the lateral size) so that one
can see the dependence 
on the source width only.
It is clearly seen that for a small source width the quality factor drops significantly. From the data shown in Fig. 4(a) and 4(b) it can therefore be argued that the finiteness of the source or the system's latteral size or both give 2-D characteristics in what would essentially be a 1-D resonance for the infinite system and result in a reduced
quality factor. By an infinite system we mean infinite latteral size and that the incoming waves have plane wave front. In our system when the latteral size is small the resonance is supported by a smaller number of scatterers and is forced to terminate at the edges. Also, when the system is large but the source is small the resonace occupies a smaller fraction of the cavity area and has magnitude that decreases as one moves away from the center of the resonance. The resonance can decay in time not only along the propagation direction y as the true 1-D resonance does but also along the latteral direction. In both cases (small latteral width or small source) the result would be a smaller quality factor. Now, for the case when both the system and the source is very large the resonance can be extended in the cavity region and be very close to the 1-D resonance. As seen in Fig. 4(b) for source width equal to 30a and  $L_{w}=40a$ the quality factor is $\sim 174$. That value is close to $\sim 180$ which is the value of the quality factor as obtained from the FDTD that models the infinite system (Bloch boundary conditions) (dotted horizontal line in Fig. 4). The solid horizontal lines in  Fig. 4(a) and 4(b) give the quality factor for the infinite system as obtained from the TMM. The respective value is $\sim$230. We have also looked at the dependence of the quality factor on the latteral width and kept the width of the source constant and equal to 2a. What we found was that the quality factor increases as the latteral width increases, reaches a maximum value , then startes to decrease and eventually saturates. This is the result of a combined effect. As described previously when the latteral width is small this causes the resonant peak to terminate at the edges. As the latteral width increases the resonance occupies the maximum area it can --according to the specific source width. Further increasing the latteral width will not cause the resonance to occupy more area of the cavity along the latteral direction x. It only allows it to decay in time along the x direction as well and therefore the quality factor starts to lessen. Another component  that can cause the quality factor to decrease with decreasing source width is the angle span of the source that increases with decreasing source width. This would cause the peak to shift and broaden (towards the higher frequencies). We have numerically checked that the magnitude for the off normal components of the incoming wave is relatively small. Also, we have not observed significant shift in frequency with decreasing source width in the FDTD results. We have observed though that the height of the transmission peak decreases as the source width decreases. It approaches the value of one as the source width approaches the lateral size of the system (for the large system $L_{w}=40a$).  It becomes one for the infinite system modeled either through the TMM or the FDTD with Bloch boundary conditions in the lateral direction of the crystal.    
   
Fig. 5(a) shows the transmission calculated with the TMM method versus the reduced frequency u=a/$\lambda$ for a 1-D cavity of width $L_{c}$/a=0.985 and $N_{c}$=4, for different imaginary parts in the air dielectric constant. Introducing 
imaginary parts  with values of 0.02-0.05 in the air dielectric constant in the TMM system, has been suggested as a mechanism to model out of plane losses in \cite{rattier}. This leads as well to a reduced Q factor value. $N_{c}$ is the number of rows of air cylinders in each side of the 1-D dielectric cavity. As expected the transmission peak decreases with Im $\epsilon$ while the width of the transmission increases with Im $\epsilon$.
We have also examined how the size of the system along the propagation direction affects the characteristics of the resonance.
In Fig. 5(b) the dimesionless linewidth 
$\Delta$/a
- which is defined as $\delta\lambda$/a  (where $\delta\lambda$ is the width in wavelength of the 
resonance at the half maximum value) - is plotted versus $N_{c}$ (the number of rows of cylinders on
each side of the dielectric cavity) in semi-logarithmic scale. As mentioned previously $N_{c}$ is related with the thickness of the system $L_{th}$ through the relation $ L_{th}=2N_{c}b+L_{c}$ ($L_{th}$ is shown in Fig. 1). 
 A linear relation in the semi-log scale between
$\Delta$/a with $N_{c}$ is obtained, meaning that the quality factor Q, which is inversely propotional to $\Delta$/a would increase exponentially 
with $N_{c}$. That is consistent with results for other defect
cavities \cite{SigalasPRB3815,sedp411}. In Sigalas et. al. \cite{SigalasPRB3815}, where an air defect in a dielectric array is studied, it was shown that when absorption 
is introduced in the dielectric the quality factor saturates and does not increase any further as
the size of the system increases. Such saturation with the introduction of an imaginary part  
in the dielectric of the holes 
is observed in our case also,
as can be seen in Fig. 5(b).  
We also observe (as in \cite{SigalasPRB3815}) that the saturation value of the linewidth $\Delta$ is smaller the larger the imaginary part in the dielectric.
  
In Fig. 6 we present the results achieved with the TMM of the transmision height versus $N_{c}$ for 1-D cavity of width $L_{c}$/a=0.985 for three different values of Im $\epsilon$ of the air dielectric. Notice that for Im $\epsilon$=0 the height is always one, i.e perfect transmision. However for nonzero Im $\epsilon$, the transmission height drops as $N_{c}$ increases. Two experimental values are shown for almost the same width of $L_{c}$/a as the one in the calculations. Notice that this suggests that the 
Im$\epsilon$ that can fit the experimental data \cite{rattier} can have a value close to 0.04. 

For the ideal case with no losses present the quality factor was calculated by the TMM for various
cavity widths $L_{c}$ that led to peaks that span most of the bandgap.
In Fig. 7 the quality factor is plotted versus the reduced frequency of the peak. (Peaks
too close to the edges of the bandgap were not used because the determination of their Q  
would be vague as stated earlier in this paper).
For every frequency in the plot there are three peaks that are
characterized by different  orders and correspond to different $L_{c}$ values (see Fig. 2) . So three
different curves are recovered when grouping the peaks according to order. Higher order peaks are
characterized by a higher value of the quality factor. For every order, the quality factor is maximum
for  a 
frequency close to the center of the band gap and reduces as the frequency approaches either
edge of the band gap. 
The calculations above were performed with a grid lattice dividing the lattice
constant a into ten intervals. We have investigated the dependence of the quality factor with the 
accuracy of the TMM given by the number N of intervals the lattice constant is divided to, i.e. N is the number of grid points.
It was done for three different peaks. One close to the band gap center and the others close
to the lower and higher frequency band edge respectively. It is seen from Fig. 8(a) that the quality
factor increases with increasing accuracy and eventually saturates at a value. The same holds for the quality factor calculated from the FDTD method. This is shown
on Fig. 8(b) for cavity spacer value Lc/b=1. The FDTD data of Fig. 8(b) are obtained with Bloch boundary conditions applied in the lateral (x) direction. This way one has an infinite source with a plane wave front as in the TMM. The saturation values of the Q-factor for the TMM and FDTD are $\sim$230 and $\sim$250 respectively. This difference might be due the different way of calculating the transmission coefficient in the two methods as was mentioned before in Section III. 

\vspace{1cm}
{\bf\large IV. Conclusions} 
\vspace{0.5cm}

We have studied the properties of 1-D dielectric cavity structure in a 2-D hexagonal array with
the transfer matrix method and the finite difference time domain method.
Both methods agree  in the position of the defect peak and both yield good agreement (for the peak's position) with the experiment \cite{rattier} as well. The quality factor though shows sensitivity to a lot of parameters such as the size of the system (both lateral
and along the propagation direction), the type of the incoming EM fields and losses out of the plane of periodicity. In that context the two numerical methods (FDTD and TMM) were compared.

\vspace{1cm}
{\bf\large Acknowledgments}
\vspace{0.5cm}

The authors would like to thank Mario Agio for providing the FDTD code and helpful discussions. We would also like to acknowledge H. Benisty, M. Rattier, and M. Sigalas for
helpful discussions. Ames Laboratory is operated by the U.S Department of Energy by Iowa State University under Contract No. W-7405-Eng-82. This work is supported by the IST project PCIC of the European Union and the NSF grant INT-0001236.

\newpage 

\begin{figure}
{\includegraphics[width=15cm,angle=0]{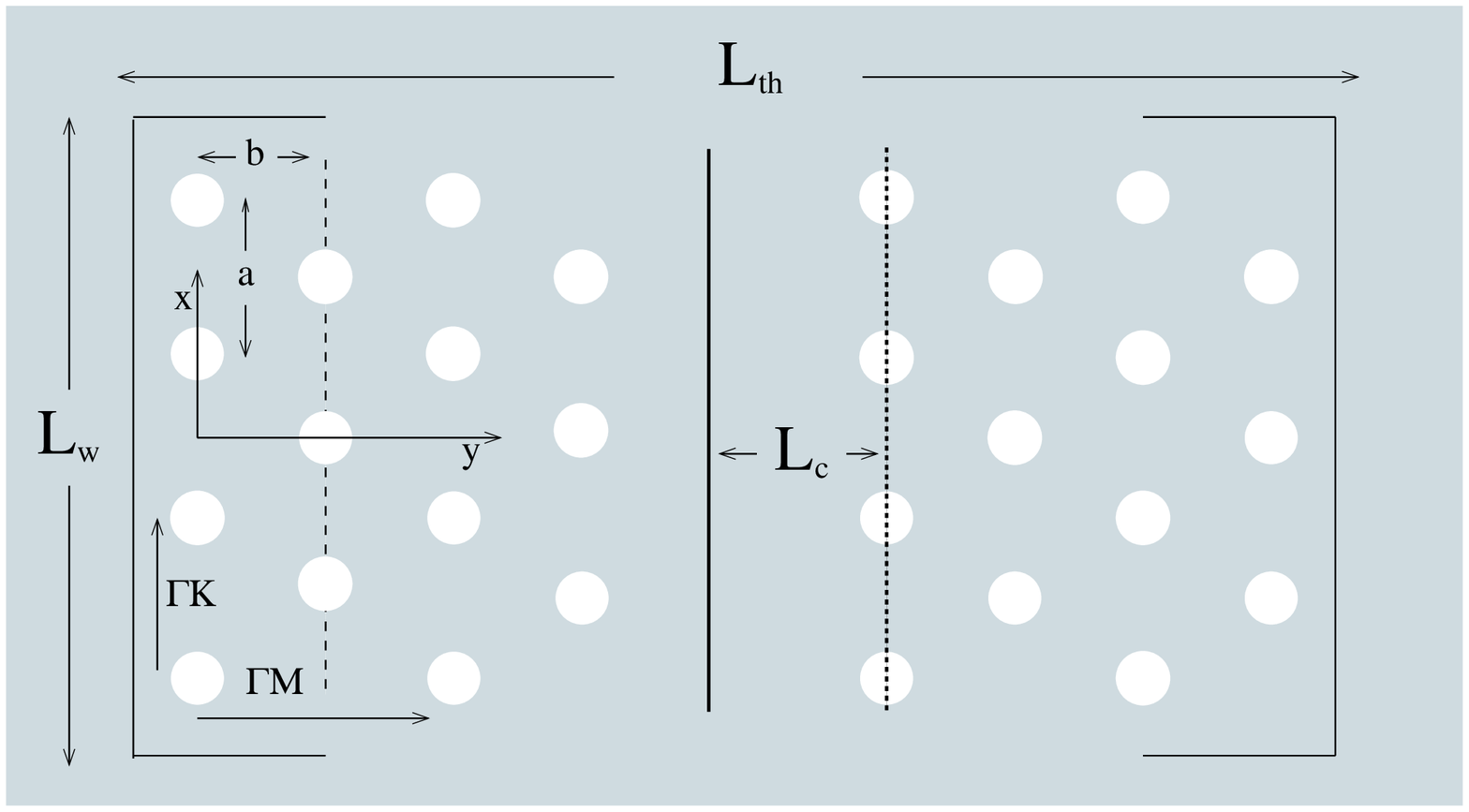}}

\caption{The cavity structure under study. The bulk crystals symmetry directions $\Gamma$M
and $\Gamma$K are shown. $ L_{c}$  refers to the cavity width. For the actual crystal, 2b=$\sqrt3$a,  but in the  TMM  due to necessary  approximations,  2b=1.7a. $ L_{w}$ is the lateral width of the cavity while $ L_{th}$ is the thickness of the cavity system.} 
\end{figure}

\begin{figure} 

{\includegraphics[width=15cm,angle=0]{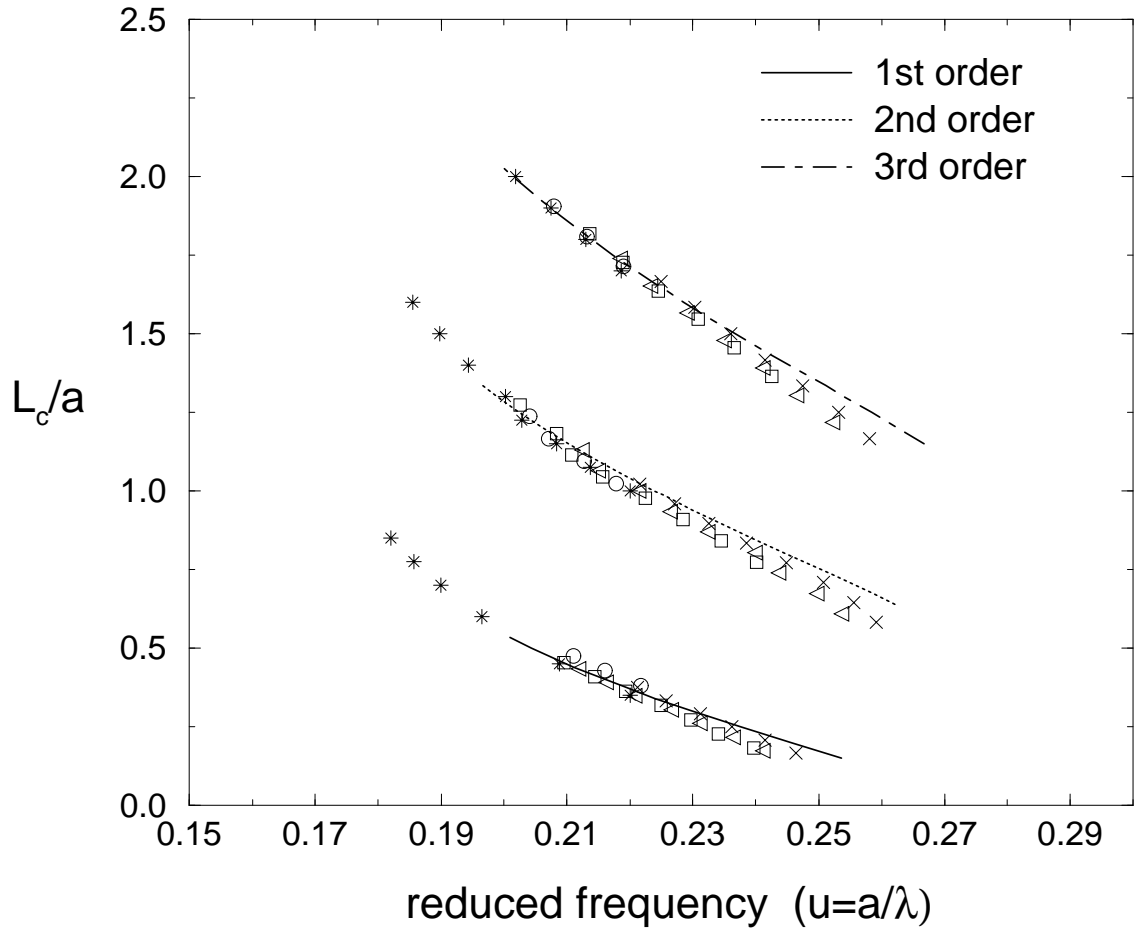}}

\caption{The cavity width versus the reduced frequency of the corresponding observed cavity resonant
peaks. The results are obtained with the TMM (solid dotted and dot-dashed line). The experimental results for various lattice constants a are shown for comparioson. The stars, circles, squares, triangles (left), and x are for a=200, 210, 220, 230, 240 nm respectively. }
\end{figure}

\begin{figure}

{\includegraphics[width=15cm,angle=0]{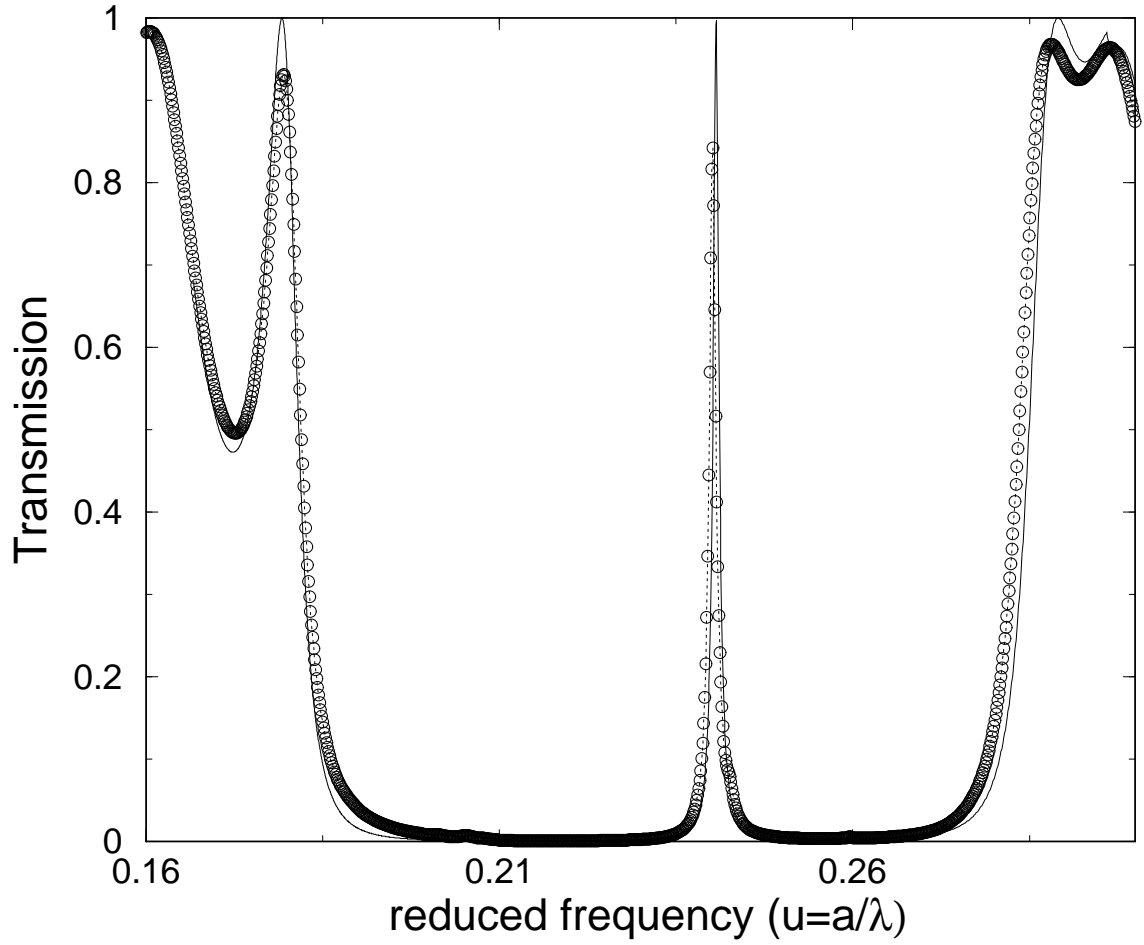}}

\caption{Comparison between transmission results obtained with the transfer matrix technique (solid line)
and the FDTD (dotted line with circles), for $L_{c}$/a=0.85. In order to make the comparison in this particular case we considered
an approximate triangular structure (2b=1.7a) in the FDTD  method too.}
\end{figure}

\begin{figure}
{\includegraphics[width=12cm,angle=0]{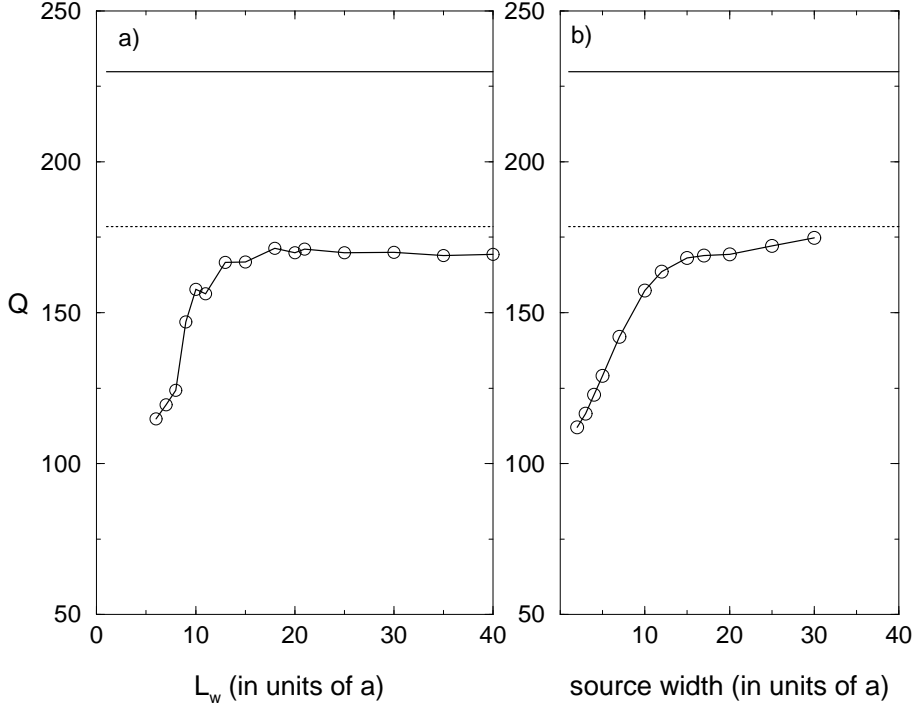}} 
\caption{a) Quality factor Q for a structure with a width $L_{c}$/b= 1. (2b=$\sqrt3$a is taken)
calculated with the FDTD versus $ L_{w}$ (in units of a) (see Fig. 1). The source is 
choosen to have length approximately half of the latteral length of the structure for all $L_{w}$. 
b) Quality factor Q calculated with the FDTD (solid line with circles) with the same $L_{c}$/b as in (a), versus the source width (in units of a). $L_{w}$ is kept constant and equal to 40a. 
 In both (a) and (b) the bold dotted line represents the value of the quality 
factor calculated with the same method but with periodic (Bloch) boundary 
conditions along the lateral direction (i.e. perpendicularly to the 
propagation direction). The source used in the calculations corresponding 
to that case is a pulse but with plane wave front. 
The horizontal solid line is the TMM result.}    
\end{figure}

\begin{figure}

{\includegraphics[width=15cm,angle=0]{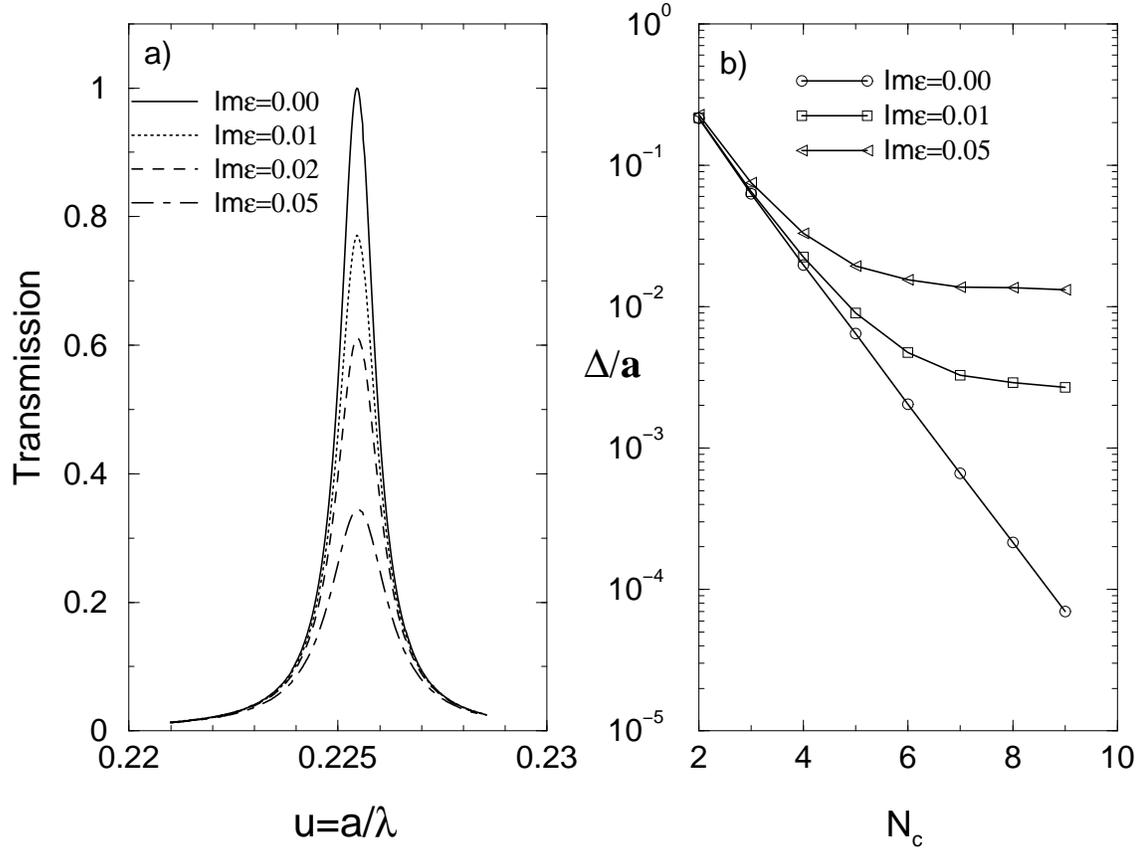}}

\caption{a) Transmission calculated by the TMM versus reduced frequency 
u=a/$\lambda$ for different values of Im$\epsilon$ of the air dielectric. 
$L_{c}/a=0.985$ on the cavity  resonance (for $N_{c}=4$). b) Dimensionless linewidth ($\Delta$ is  $\delta\lambda$ i.e. the wavelength 
width that corresponds to transmission half of the peak's maximum)  versus $N_{c}$ (number of rows in each side of the cavity) for the lossless case as
well as the cases with non-zero imaginary part in the air dielectric constant.}
\end{figure}

\begin{figure}

{\includegraphics[width=14cm,angle=0]{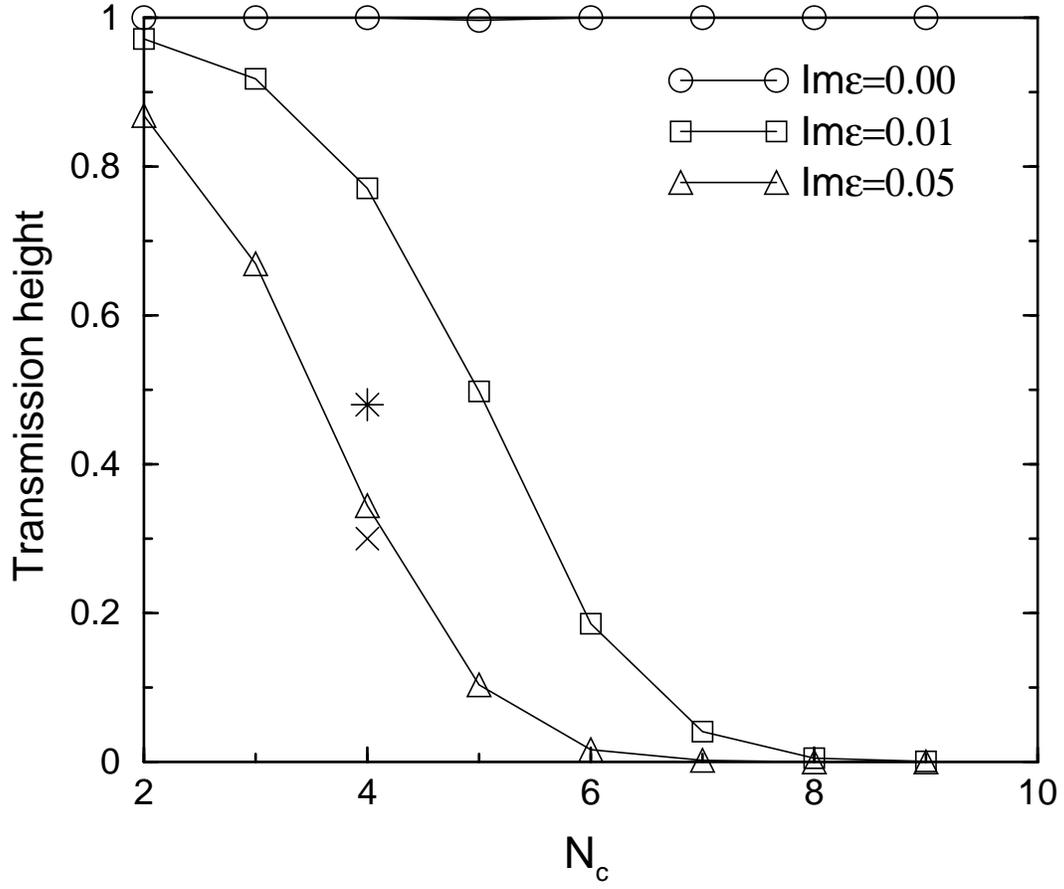}}

\caption{Transmission height of the resonance for $L_{c}$/a=0.985 for three values of the imaginary part of the air dielectric constant, Im$\epsilon$. Two 
experimental values for the cavity with $N_{c}$=4 are shown for comparison. 
The star and x correspond to a=230 nm with  $L_{c}$/a=1.07 and a=220 nm 
with  $L_{c}$/a=0.98, respectively.} 
\end{figure}

\begin{figure}  

{\includegraphics[width=15cm,angle=0]{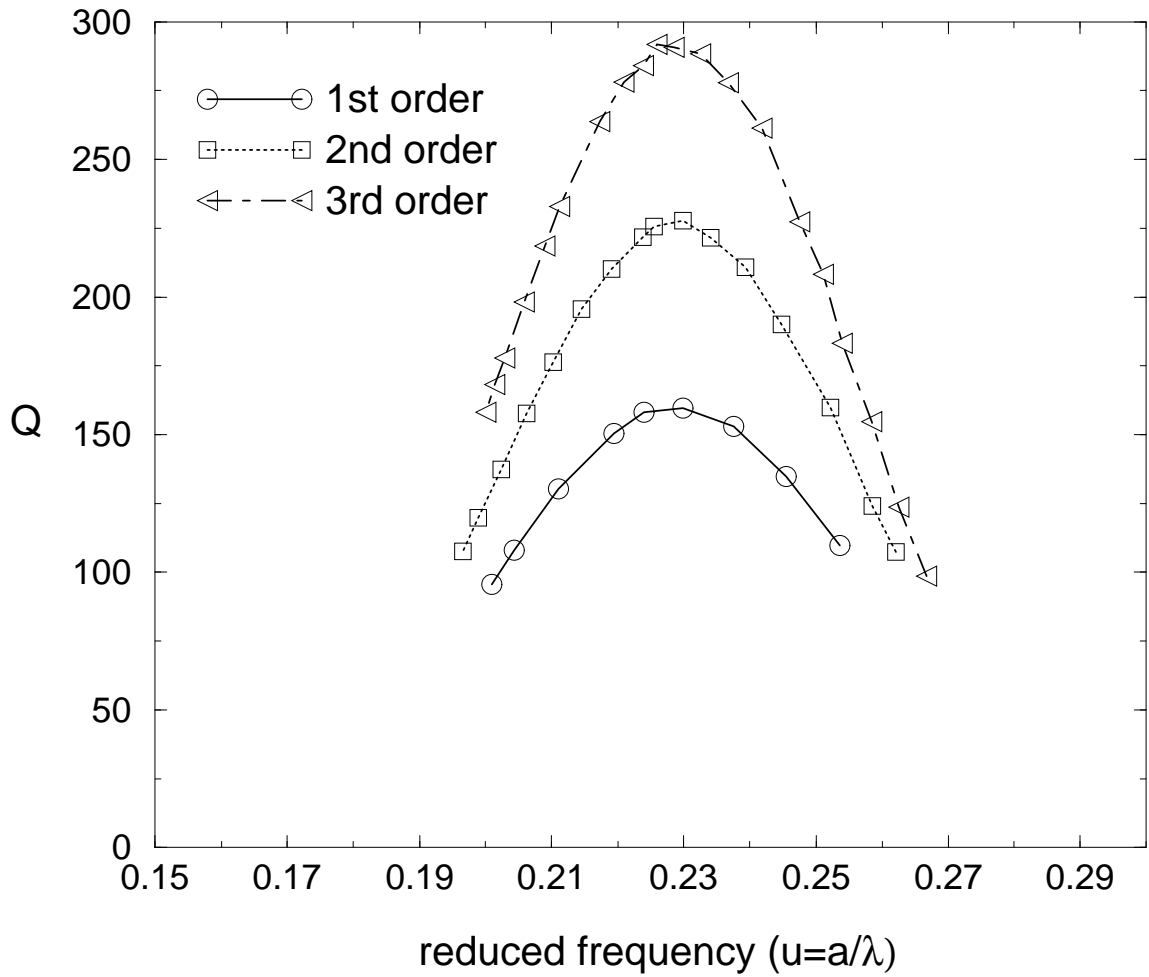}}

\caption{Quality factor calculated with the TMM
versus the reduced frequency for first-, second- and third- order resonant peaks.}
\end{figure}

\begin{figure}

{\includegraphics[width=14cm,angle=0]{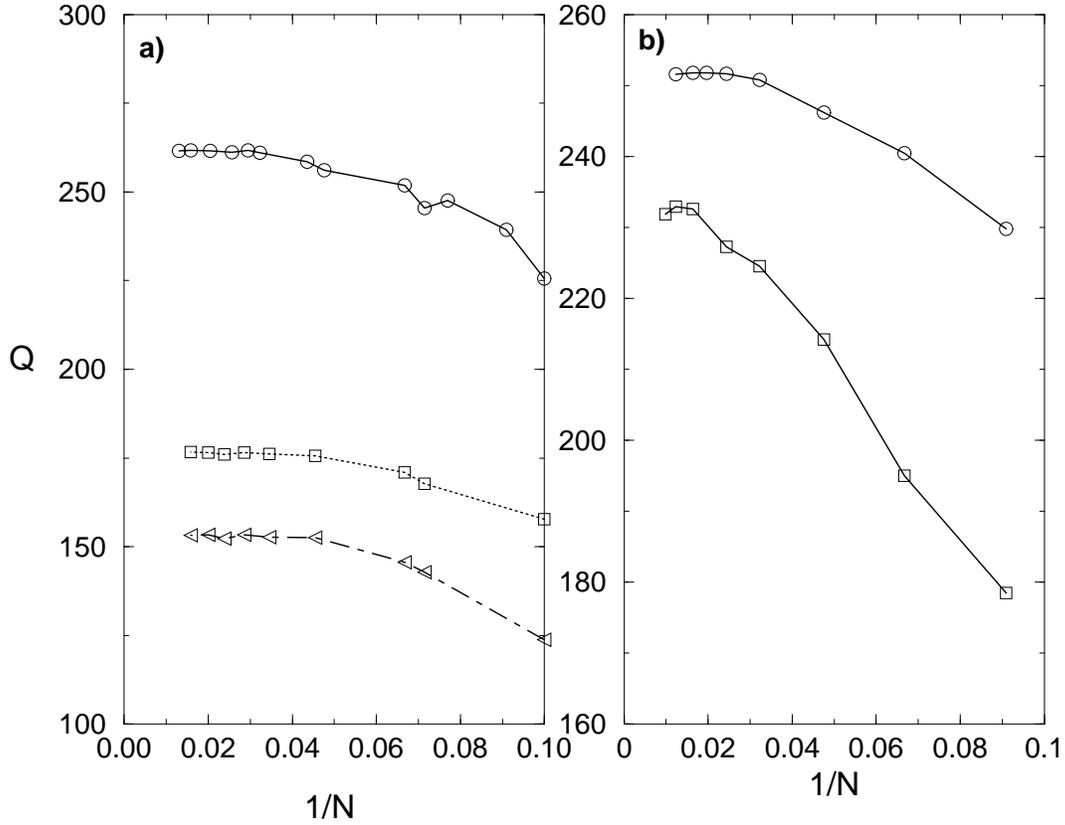}}

\caption{a) Quality factor  for three different cavity peaks with  reduced frequencies u=0.2255 (solid line), u=0.2063 (dotted line) and u=0.2628 (dotted-dashed line)
versus 1/N  where N is the number of grid spacings that
the lattice constant a is divided into in the computation. The peaks at 
u=0.2255 and u=0.2063 are second order peaks and have $L_{c}$/a=0.985 
and $L_{c}$/a=1.2  respectively. The peak at u=0.2628 is of third order 
(with $L_{c}$/a=1.2). The quality factor saturates at a value as the 
numerical accuracy of the method increases. b) Quality factor for Lc/b=1. as a function of the grid accuracy 
calculated from the FDTD method with Bloch boundary conditions across 
the latteral direction (solid line with squares)  and from the TMM (solid line with circles).}

\end{figure}
  
\end{document}